# A next generation Ultra-Fast Flash Observatory (UFFO-100) for IR/optical observations of the rise phase of gamma-ray bursts

B. Grossan[*,a,b,c], I. H. Park[d], S. Ahmad[e], K. B. Ahn[f], P. Barrillon[e], S. Brandt[g], C. Budtz-Jørgensen[g], A. J. Castro-Tirado[h], P. Chen[i], H. S. Choi[g], Y. J. Choi[j], P. Connell[k], S. Dagoret-Campagne[e], C. De La Taille[e], C. Eyles[k], I. Hermann[j], M.–H. A. Huang[l], A. Jung[d], S. Jeong[d], J. E. Kim[d], M. Kim[d], S.-W. Kim[f], Y. W. Kim[d], J. Lee[d], H. Lim[c], E. V. Linder[c], T.–C. Liu[i], N. Lund[g], K. W. Min[j], G. W. Na[d], J. W. Nam[i], M. I. Panasyuk[b], J. Ripa[d], V. Reglero[k], J. M. Rodrigo[k], G. F. Smoot[b,c,a], J. E. Suh[d], S. Svertilov[b], N. Vedenkin[b], M. –Z. Wang[i], I. Yashin[m], M. H. Zhao[d]; [a]University of California, Berkeley, USA; [b]Extreme Universe Laboratory, Moscow State University, Russia; [c]Institute for the Early Universe, Ewha Womans University, Seoul, Korea; [d]Research Center for MEMS Space Telescope, Ewha Womans University, Seoul, Korea; [e]University of Paris-Sud 11, Orsay, France; [f]Yonsei University, Seoul, Korea; [g]Technical University of Denmark, Copenhagen, Denmark; [h]Instituto de Astrofísica de Andalucía - CSIC, Granada, Spain; [i]National Taiwan University, Taipei, Taiwan; [j]Korea Advanced Institute of Science and Technology, Daejeon, Korea; [k]University of Valencia, Valencia, Spain; [l]National United University, Miao-Li, Taiwan; [m]Moscow State University, Russia

## ABSTRACT

The *Swift* Gamma-ray Burst (GRB) observatory responds to GRB triggers with optical observations in ~ 100 s, but cannot respond faster than ~ 60 s. While some rapid-response ground-based telescopes have responded quickly, the number of sub-60 s detections remains small. In approximately mid-2013, the Ultra-Fast Flash Observatory-Pathfinder is expected to be launched on the *Lomonosov* spacecraft to investigate early optical GRB emission. Though possessing unique capability for optical rapid-response, this pathfinder mission is necessarily limited in sensitivity and event rate; here we discuss the next generation of rapid-response space observatory instruments. We list science topics motivating our instruments, those that require rapid optical-IR GRB response, including: A survey of GRB rise shapes/times, measurements of optical bulk Lorentz factors, investigation of magnetic dominated (vs. non-magnetic) jet models, internal vs. external shock origin of prompt optical emission, the use of GRBs for cosmology, and dust evaporation in the GRB environment. We also address the impacts of the characteristics of GRB observing on our instrument and observatory design. We describe our instrument designs and choices for a next generation space observatory as a second instrument on a low-earth orbit spacecraft, with a 120 kg instrument mass budget. Restricted to relatively modest mass, power, and launch resources, we find that a coded mask X-ray camera with 1024 cm$^2$ of detector area could rapidly locate about 64 GRB triggers/year. Responding to the locations from the X-ray camera, a 30 cm aperture telescope with a beam-steering system for rapid (~ 1 s) response and a near-IR camera should detect ~ 29 GRB, given *Swift* GRB properties. The additional optical camera would permit the measurement of a broadband optical-IR slope, allowing better characterization of the emission, and dynamic measurement of dust extinction at the source, for the first time.

## 1. INTRODUCTION

Gamma-ray burst (GRB) studies entered a renaissance when it became possible to image the GRB afterglows with ~ arc sec resolution. Such observations enable investigations of environment, distance, location within galaxies, characteristics of hosts, extinction, and a plethora of other topics impossible with the degree-scale positions from

*Bruce_Grossa@lbl.gov

gamma-ray instruments alone. Since the discovery of GRB optical afterglow from a *Beppo-Sax*[1] GRB position, the *Swift* observatory has become a prolific source of optical data, often providing the earliest measurements of hundreds of GRB. The critical changes in instrumentation that made optical measurements possible include the use of wide-field coded-mask X-ray cameras which detect and locate GRBs to within a few arc min (instead of the previous deg. scale positions) and trigger other instruments. On the *Swift* observatory, the Burst Alert Telescope (BAT) detects the initial burst of emission (called prompt emission) in the gamma- to X-ray bands. *Swift* then re-orients the spacecraft to point the UVOT UV-optical telescope and the XRT focused X-ray telescope for high-resolution imaging, producing ~arc sec positions for multi-band follow-up. The typical response time for optical imaging is about 100 s after a gamma/X-ray trigger, with a minimum of about 60 s (Fig. 1).

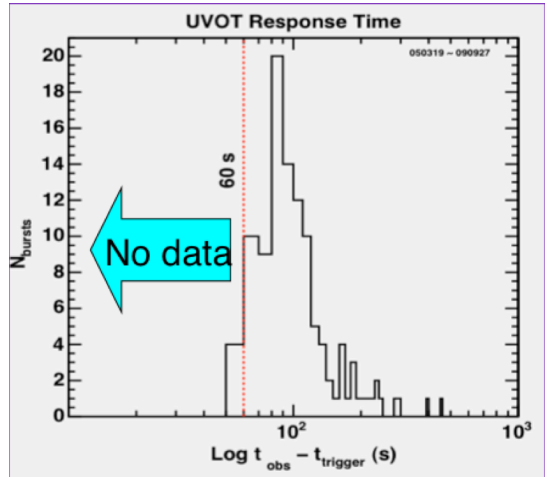

**Fig. 1** Histogram of *Swift* Optical Response Time

In contrast to the rich data in the post- 60s regime, there is no similar data base of optical or IR measurements in the sub-60 s regime, and almost nothing is known about the rise phase of optical/IR emission in this regime. There are a few ground-based optical telescopes purpose-built to rapidly respond to GRB triggers from space instruments, with very important contributions (e.g. the ROTSE telescopes[23] and TORTORA[2]). However, only ~ a dozen good-quality light curves have been produced in the sub-60 s regime, and few rise times have been measured below 60 s. Challenges of weather, limited overlap with the field of view (FOV) of spacecraft, daytime background, and loss of sensitivity due to atmospheric background have proved to be formidable challenges for the ground-based experiments. New instrumentation is therefore required to make progress in acquiring rapid –response (sub-60 s) optical and IR data.

In this paper we use typical literature terminology for a *Swift*-like instrument, implicitly allowing for exceptions and unusual cases. The typical long GRBs produce a very bright burst of "prompt" emission in the gamma-ray to X-ray bands, for 2- ~60 s, with wide variation in duration and structure. After this time, gamma and X-ray emission are typically observed to drop very quickly, with later X-ray emission continuing in a power law time decay, the "afterglow" phase. Optical emission observed during the bright gamma-ray emission is also referred to as "prompt" emission. The long-term behavior of the optical-IR emission also shows power law decays on the hours-days time scales. An obvious transition between bright, rapidly variable optical prompt emission and power law decay optical emission has been measured in the "Naked Eye" burst 080913b[3]; we assume such distinct prompt and afterglow emission components here.

In the remainder of this paper, we outline the science questions and characteristics of GRBs which directly motivate our rapid-response optical/IR instruments, briefly describe our current efforts to measure GRB optical emission in the sub-60 s regime, and outline our designs and plans for a more ambitious next-generation rapid-response space mission and instruments.

## 2. RAPID-RESPONSE INSTRUMENT MOTIVATION

Scientific questions on the nature of GRBs drive the requirements for our instruments; we therefore briefly describe some principle science questions requiring rapid (sub-60 s) response and applicable measurements.

### a. Emission & jet physics

Sari & Piran (1999)[4] made a set of predictions of early optical emission which give strong constraints on the emission mechanism, via: (1) Measurement of a delayed optical emission peak compared to the gamma-ray peak; this would provide evidence of emission from internal shocks in a relativistic jet. (2) Measurement of an early and bright optical peak during the gamma-ray emission, up to 15th mag; such a measurement is predicted for reverse shock emission. (3) Measurement of no early and bright optical peak; this measurement provides evidence of a magnetically dominated (magnetic energy >> baryon rest mass energy) jet and burst emission mechanism. Rapid-response measurement of the brightness and time scale of the early (usually sub-60 s) optical emission is therefore a critical determinant of emission mechanisms of GRBs.

Measurement of polarization in early GRB emission is also important to the study of emission mechanisms, as significant polarization can be a "signature" of synchrotron origin. Synchrotron emission by shocked electrons is the basis of most early emission models, therefore polarization measurements are a means to verify (or refute) these models. Such measurements have not yet been made (polarization has only been measured in late time emission, due to interaction with the interstellar medium, not due to prompt emission processes).

Measurements of rapid variability in prompt gamma-ray emission identify the origin as internal-shock emission (e.g. [5]). Optical rapid-response and high time resolution are therefore required to investigate the role of the internal shock mechanism in optical emission in the same way.

Measurement of the bulk Lorentz factor (BLF) in the GRB jet is critical in understanding jet physics. A simple, nearly model-independent argument by Molinari et al. (2007)[6] shows that the BLF can be measured from the time of the early optical emission peak; a rapid-response instrument fast enough to catch the optical emission peak is therefore required. Measurement of a correlation (or not) with gamma-ray measured BLFs, would support (or not) the same emission mechanism and location for both. A separate optical BLF measurement would allow comparison with the gamma-ray BLF, a new science capability for probing these mechanisms.

### b. Correlations of prompt optical emission and other observables

Measurements of sub-60 s optical light curves would add information to a variety of GRB observable correlation studies. Some such studies search for an independent predictor of luminosity and/or distance to enable the use of GRBs as cosmological probes. (Instrumental biases[7],[8] are important in such studies, and recent work considering multi-parameter adjusted correlations should be considered[9].) Little work has been done relating sub-60 s optical emission properties to other GRB observables. Panaitescu & Vestrand (2008[10]) published an extensive survey of the earliest optical observations available at the time, and showed that optical emission at $t=10^3$ s correlates with the rise time of the optical emission (Fig. 2a), potentially useful to the study of cosmology at the great distances of GRBs. However, this correlation result lacks data: only 5-11 of the 30 GRB in the sample had fast enough response to measure the rise time (Fig. 2b). To test this, and other early optical emission properties rigorously, many more rapid-response measurements are required.

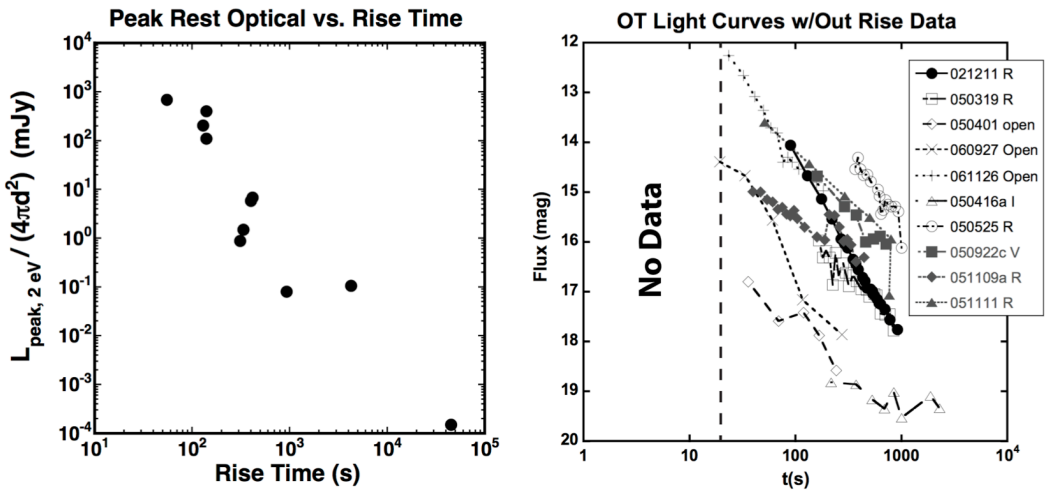

**Fig. 2a**, left, shows the correlation of peak optical emission and rise time, from the data given in Panaitescu & Vestrand (2008; [10]). **Fig. 2b**, right, shows that the *majority* of the light curves studied in Panaitescu & Vestrand (2008) do not have measured rise times; the correlation cannot be reasonably tested until more rapid-response measurements produce a more complete data base of early rise times.

### c. Multi-messenger signals

Observations are now routinely made of non-photon signals predicted for GRBs: high and ultra-high energy cosmic rays (e.g. ARGO-YBJ[11]), neutrinos (e.g. ICECUBE[12]), and gravitational waves[13]. Fast-response optical observations can test, e.g., Lorentz violations, from the time delay between different energy photons, or between photons and non-photon emission. Such observations would revolutionize astronomy and greatly improve our understanding of black holes, neutron stars, cosmology (e.g. [14]), and strong field gravity. Fast-response optical capability is required to associate the prompt photon, particle, and gravitational wave signals with that from prompt optical emission.

### d. Time-evolution of local dust

Measurement of rapid evolution of the optical-IR slope provides rich information on processes and environment local to the GRB. GRBs are associated with massive stars, which are associated with large dust and gas columns. Observations of X-ray afterglows provide evidence of gas absorption columns for most bursts (equivalent $N_H \sim 10^{22}$ cm$^{-2}$), normally associated with $A_V > 1$-10 mag (e.g., [15,16,17]). However, typical IR-UV observations do not show such large extinction columns (e.g. [18]). This discrepancy may result from the very-rapid destruction of the circumburst dust by a prompt optical-UV flash (e.g. [19,20]). If this process occurs, rapid early-time color and brightness evolution would be observed as the radiation "burns" its way through the dust, changing from extremely red to blue with the brightening of the optical emission. Direct detection of this process would be a major result, opening new avenues for studying the GRB environments and progenitors. Current observations are too slow measure this phenomenon, as dust destruction should happen within seconds of the GRB, with minimal change after t > 60 sec, again requiring rapid-response to make this measurement.

## 3. THE ULTRA-FAST FLASH OBSERVATORY PROGRAM AND PATHFINDER

### a. The UFFO concept

The Ultra-Fast Flash Observatory (UFFO) concept is broadly like that of *Swift*, in that a coded-mask X-ray camera locates GRBs in real time, to within a few arc min, so that more narrow FOV instruments may then observe the GRB.

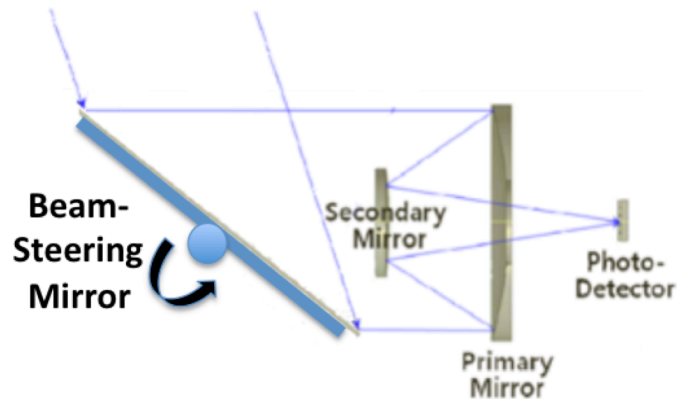

Fig. 3 The UFFO concept incorporates a low-mass/moment of inertia steering mirror which may be re-pointed more rapidly than an entire spacecraft or telescope structure.

The UFFO concept aims to accomplish what *Swift* cannot, that is, to start optical/IR observations of the GRB much earlier. The concept has two novel elements: First, UFFO uses a beam-steering mirror. The 60 s hardware response time limit of *Swift* is caused by the finite time required to re-orient the *Swift* spacecraft and the associated settling time.

Instead of re-orienting an entire spacecraft, why not move only a light mirror, for much faster response? The UFFO has a simple, flat beam-steering mirror in front of its optical telescope (Fig. 3). Using modern materials and engineering techniques, such mirrors can be constructed to be much less massive and with lower moment of inertia than an entire spacecraft (or telescope structure), for much faster movement and settling.

Second, unless we obtain a very large spacecraft, mass allotment, and other resources, we do not plan a focused X-ray telescope. This is a distinct strategy from that of *Swift*, which responds to GRB triggers with both pointed UVOT optical and XRT X-ray instruments. The XRT produces arc sec quality positions for even highly optically extinguished GRB and measurements of the X-ray afterglow, critical for afterglow science. However, the XRT is a complex, expensive instrument, necessarily large due to the weak converging power of X-ray mirrors. Such an instrument would not be possible as part of a relatively simple, low-mass and low-cost mission envisioned for a modest UFFO program. Consider, however, that almost any optical or IR instrument detecting a GRB would yield an arc sec scale position suitable for follow-up work. Further, lack of a pointed X-ray telescope pre-selects our desired GRBs, those with bright optical/IR emission that might be detected in short, high-time resolution exposures, those which would be good candidates for ground-based follow-up spectroscopy, imaging during later, faint epochs, etc. In addition, another space observatory with a focused X-ray telescope could be used to obtain X-ray afterglow information, using the Gamma-Ray Burst Coordinates Network (GCN)[21] to communicate the GRB position. Several such instruments are expected to be in operation in the near and medium term.

One more unusual element of the UFFO program, again for the case of constrained mass budgets, is our modest size requirements for the X-ray coded mask camera trigger. We are asking new questions about the already well-studied population of GRBs observed with *Swift* and other instruments; studying only the brightest bursts of this known population is sufficient, as long as reasonable numbers of measurements result. The extraordinary brightness of GRBs makes this modest instrument requirement possible: GRBs are so bright that the detection rate decreases only slowly with decreased instrument sensitivity (Fig. 4). For more typical types of sources, the detection rate decreases steeply, roughly as the –3/2 power of the measured flux; GRBs are so bright that the number of observed sources are truncated by a finite observable universe[22], *not by detector sensitivity* (i.e. $<V/V_{max}> < 0.5$) for instruments such as *Swift*. Since sensitivity falls off slower than linearly with detector area, and since GRB detection rates are relatively weakly sensitive to limiting flux (actually limiting fluence or peak flux for these transient sources), even a much smaller instrument than *Swift*-BAT can detect dozens of GRB per year.

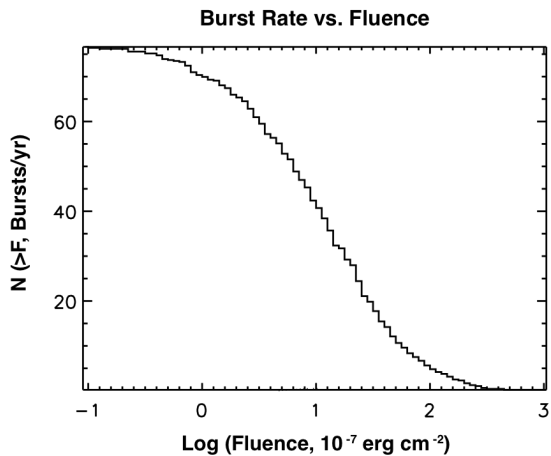

**Fig. 4** *Swift* BAT fluence histogram. From the lowest fluence burst to more than two orders of magnitude greater, the burst rate does not decrease by half. The low-fluence rolloff is due to a finite universe. (First 5 years of triggers shown.**)**

### b. The UFFO Pathfinder

To begin our study of sub-60 s optical emission, our collaboration designed and built the UFFO-Pathfinder (UFFO-P), a small proof-of-concept experiment with 191 $cm^2$ X-ray collecting area, and a 10 cm aperture optical telescope. The instrument will fly on the Lomonosov spacecraft, expected to launch in 2013 June . The beam steering mirror is mounted on a standard gimbal, and driven by stepper motors. The settle time of this system is only 120 ms, and any point in the FOV of the X-ray camera can be reached in less than 1 s. The technical details of the observatory are described in detail in other papers in this volume. The collecting area of the UFFO-P is small, but we estimate that with full operation after launch, the X-ray camera will detect about 40 GRB per year and the optical instrument will detect ~ 5 of these each year; even though this is a "pathfinder" experiment, it has significant scientific potential.

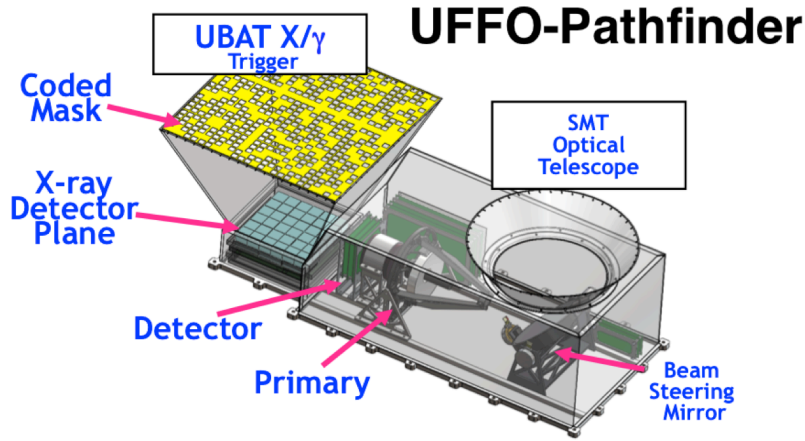

**Fig. 5** Schematic of UFFO-Pathfinder. The two main instruments are the UBAT X-ray coded mask camera, for burst detection and coarse localization, and the SMT telescope and optical camera for early measurement of optical emission. The mirror can point and settle in less than one second.

## 4. A NEXT GENERATION RAPID-RESPONSE GRB OBSERVATORY

### a. Science Goals & Design Envelope

The primary goal of our next generation instrument is to provide a statistically meaningful sample of sub-60 s optical-IR detections, dozens or more per year. There are many other worthy science goals in this field, but instruments to carry out such goals require more cost, effort, and spacecraft resources. A partial list of additional goals would include: early light curve measurements of short GRB (SGRB), optical and X-ray polarization measurements, and optical spectra. If spectra cannot be obtained, then multiple filter bands should be measured; at a minimum, broad-band spectral slopes in the optical-IR would make dynamic extinction studies (Section 2.d) possible.

The design envelope imposed by the spacecraft, launch vehicle, and launch authority greatly restricts the choices and scope of our observatory and instruments. Some of our colleagues have been informally offered "partner mission of opportunity" launches to low-earth orbit for a secondary instrument, ~ 120 kg, on a separately defined mission and spacecraft, taking place in the next few years. This sets the requirements of our proposed instruments, and we began our design apportioning roughly half of the mass for optical/IR and half for X-ray instruments.

### b. Coded Mask X-ray GRB Monitor

The *Swift* BAT X-ray camera has 5200 $cm^2$ collecting area in the CdZnTe semiconductor detector plane, sensitive to ~ 15-50 keV. Unlike typical gamma-ray detectors, this lower energy device has no "active" shielding, but rather graded high-Z material passive shielding, making for a simpler and lighter instrument. Our own coded mask X-ray GRB monitor camera is similar in design, using the same detector type, similar random mask, passive shielding design, and the same FOV (but with a square outline). The camera has 1 mm thick tungsten mask tiles, passive graded shielding (Pb-Ta-Sn-Fe) with design parameters given in the table under Fig. 7. Scaling the mass from our UFFO-P X-ray camera, we find that we can produce a 1024 cm² detecting area coded mask camera in as little as 46 kg. De-convolution of images is done on-the-fly via an on-board image processing unit shared with the other instruments (5 kg, 14 W additional).

### c. Telescope and Beam-Steering Mirror

Unlike the flux-number count relation for gamma-ray emission, optical detection of GRBs is a steep function of sensitivity, making this sensitivity critical for our science goals (Fig. 8). We calculate the mass of our telescope and beam-steering system by scaling up those from the UFFO-P, with modifications for scale. As with the UFFO-P, we chose the well-known Ritchey-Chrétien optics design, facilitating our process of modifying existing designs and using commercial parts and vendors where possible. We determined that we could produce a 30 cm aperture telescope, the same aperture size as the UVOT, with beam-steering system, within a 60 kg mass budget. We assumed that the spacecraft would be much more massive than the beam-steering mirror system, so that mirror operation would cause only negligible angular acceleration of the stabilized platform. We did, however, take note that the beam-steering mirror and structure must be able to settle with sub-second timescales, and designed to minimize mechanical oscillations.

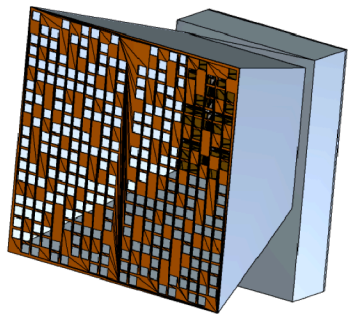

**Fig. 7** Coded Mask X-ray Camera

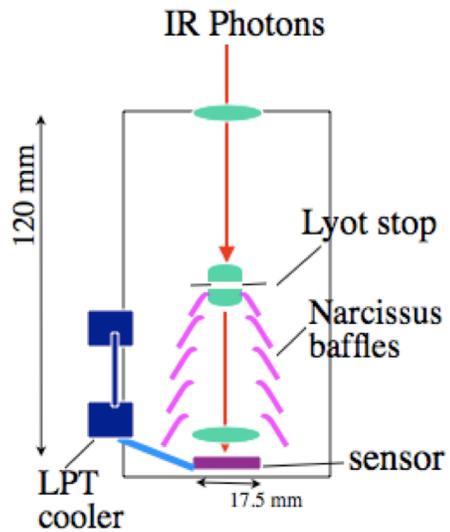

**Fig. 6** NIR Camera Block Diagram

| X-ray Camera Characteristics | | NIR Camera Characteristics | |
|---|---|---|---|
| Detector | **CdZnTe crystal** | Band | **0.9 – 1.7 µm** |
| Active Area | **1024 cm²** | Thermal Design | **Lyot Stop, Narcissus baffles, 1 cooled lens, Linear Pulse Tube Cooler** |
| Band | **15-200 keV** | | |
| Optics | **coded mask aperture** | | |
| Source Location | **r < 8.5',(90% prob., 8 σ)** | Detector | **H2RG HgCdTe array** |
| FOV | **HCFOV=1.4 Sr** | Cooling | **Linear Pulse Tube Cooler** |
| Mask-Detector Sep | **400 mm** | FOV | **17.1'** |
| Mask Width | **575 mm** | Npix | **2048 X 2048** |
| Mask Element Size | **5.5mm** | Operations modes | **500kHz read (23e-),** |
| Npix | **16,384** | (noise for 2 reads) | **100 kHz read (15 e-).** |
| Data/GRB | **<5 MB** | Data/GRB | **640 MB** |
| Mass | **46.0 kg** | Mass | **5.1 kg** |
| Power | **50.7 W** | Power | **55.4 W** |

### d. Optical & IR detectors and instruments

The *Swift* UVOT has an intensified CCD (ICCD) detector, motivated by high time resolution and wide bandwidth. These detectors use photocathodes with only ~ 10-25% quantum efficiency, which is worst to the red of 6000Å. GRB have more photons toward the red, due to their relatively steep log slope in the optical-IR, ~ –0.75 for early emission[23]. HgCdTe array detectors are sensitive from roughly 5500Å to as far red as 2.5 µm where GRB are many times brighter. CCDs and HgCdTe detectors have above 92% and 88% quantum efficiency, respectively, for much of their sensitive bands. This is a substantial gain in sensitivity, assuring that with the same aperture size, our next generation instrument using these detectors will be *significantly* more sensitive than *Swift* to GRB optical-IR emission. Many GRBs are also red due to extinction (which strongly attenuates blue light but has a much weaker effect red of 1 µm); only 15/23 bursts known to be detectable in the near-IR (NIR) would be detected by the red-insensitive UVOT[24]. A NIR instrument, for similar aperture size, would therefore also have a higher detection rate per similar GRB trigger than UVOT.

GRB optical/IR spectra are known to be featureless power laws, and the wider the measurement band (neglecting instrument background), the higher the S/N ratio. We chose to use one optical and one NIR camera to measure two wide bands simultaneously. One can select a dichroic from commercial vendors to split the beam from the telescope into optical and NIR beams with a sharp transition between the two bands and very little loss. The dichroic is placed in the telescope optical path just as the beam exits the telescope; the reflected beam is directed into the NIR camera and the remaining light continues on into the optical CCD camera. CCD camera and components are readily obtained commercially, designed for good performance with only a modest thermoelectric cooling system. Commercial cameras (7 electron read noise), suitably modified for spaceflight, would detect GRB with typical –0.75 optical-IR slope with V > 21.0 mag (5 sigma significance) in a 10 s exposure, in a 0.3 – 0.9 µm band, for our 30 cm aperture and low zodiacal background.

Space platform NIR cameras, without atmospheric background, can greatly exceed ground-based camera performance. We base our camera on the H2RG Teledyne sensor and electronics[25]. This sensor requires cryogenic cooling to reduce dark current to good performance levels. We chose a linear pulse tube cooler, which requires 2.3 kg and 55W power. to conservatively maintain the sensor at 140K in operation. Thermal background from a warm, i.e. 300 K, telescope mirror will be substantial for wavelengths longer than 2.0 µm; we therefore choose the long-wavelength cutoff of our device to be 1.7 µm, effectively eliminating this background. We designed the NIR camera, shown schematically in Fig. 6, to be robust against external heat input, yet lightweight. The optics first re-image the pupil of the telescope on the Lyot stop (a common feature of many IR cameras), which restricts the exterior view of the camera to low-emissivity mirrors reflecting the sky. The last field lens before the sensor, which is cooled, redirects light reflected from the detector back out through the entrance pupil. We also use a Narcissus baffle to reduce thermal background: this consists of a series of mirrors which restrict the view of the sensor to only the telescope beam or reflections of itself, thus protecting the sensor from warm radiating surfaces. The combination of a low-background camera design, low space background, red sensitivity, and wide filter band makes for a very sensitive instrument. We estimate an equivalent V-band sensitivity to GRB emission by assuming a –0.75 optical-IR log slope, measured read and dark noise of the H2RG sensor, and measured sky background, shown in Fig. 9. In a 10 s exposure, the NIR camera will detect GRB fainter than V=21.1 mag (5 sigma significance) in our 0.9-1.7 µm band.

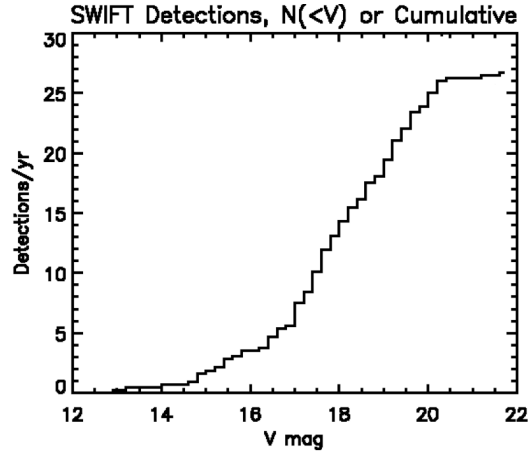

**Fig. 8** *Swift* UVOT optical (peak) flux histogram. Unlike the gamma/X-ray bands, the detection rate decreases by 4X for an order of magnitude flux increase (V=20 to V=17.5; Note also that the sharp turnover for V>20 is due to detection efficiency drop, unlike in **Fig. 4**).

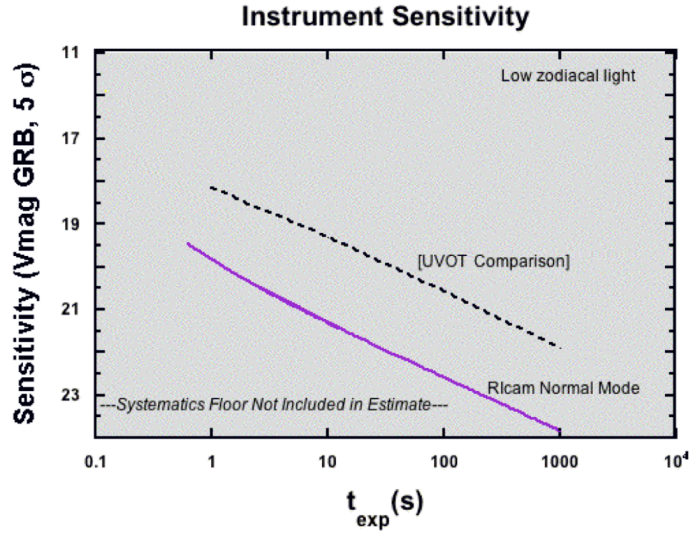

**Fig. 9** NIR camera sensitivity estimate (expressed in equivalent V-band mag for typical spectrum) for low zodiacal background. The combination of high quantum efficiency for the H2RG sensor, low IR space background, wide bandwidth, and red sensitivity make this camera significantly more sensitive than UVOT.

To illustrate the operation of simultaneous observations in two bands, we show (Fig. 10) how a simple ratio of fluxes measures changes in extinction due to GRB dust evaporation, as described in section 2.d.

### e. UFFO-100 and other configurations

We designed the UFFO-100 package (Fig. 11) using a folding mirror in order to fit within a square footprint with minimum volume, and within a 122 kg, 150 W basic budget. (The basic budget includes X-ray, NIR, optical instruments and their electronics and data communications, the optical telescope, beam-steering system, and image processing unit; all other systems are assumed provided by the spacecraft). Other configurations are possible for other launch vehicles and instruments; in order to respond to as wide a range of launch opportunities as possible, we are also designing add-on instruments and their observatory configurations.

For a small amount of mass and additional power, a new technology can bring a significant improvement in sensitivity. Since *Swift*, CdZnTe and CdTe detectors have been shown to be sensitive down to 4-5 keV, rather than *Swift*'s lower energy threshold of 15 keV[26]. For a photon flux log slope of -1.57 ([27]), typical for long GRB, this reduction in low energy threshold yields more than a factor of 2.7 increase in the number of source photons. There are two requirements to achieve this: First, the detectors must be cooled to –20 C, requiring more power and mass for the cooling system. Second, very low-noise electronics must be used, requiring experience in the design and fabrication of such systems. The substantial increase in the photon flux is a substantial reward for the additional effort, however. Design work in support of this option is in progress.

In order to verify the proposed synchrotron origin of prompt optical emission, it would be valuable to have rapid-response polarization measurements. For rapidly variable phenomenon such as GRBs, however, it is desirable to measure all polarization information simultaneously, requiring multiple detectors or cameras. For our current estimates, the additional detectors would require greater than 15 kg and 20 W. Deployment and design of the polarimeter therefore depends critically on the launch mass, volume, and power finally available. Design of an add-on polarimeter is in progress.

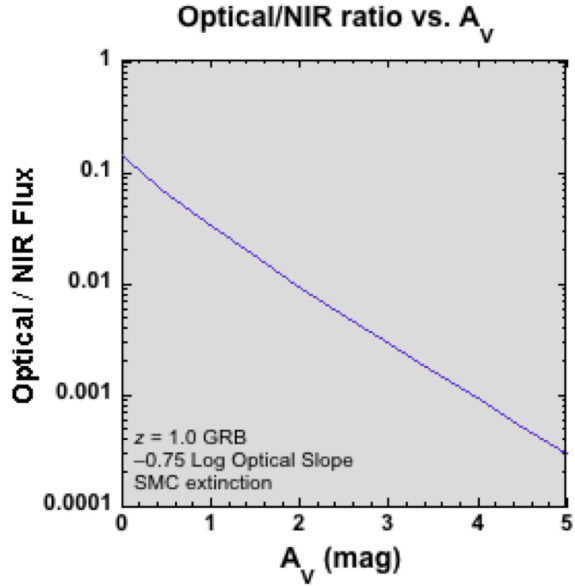

**Fig. 10** Extinction measurement via optical/NIR flux ratio. The plot shows that extinction may be monitored, even dynamically, via the ratio of flux in optical and NIR bands, (The calculation assumes a GRB at a red shift of $z$=1.0, typical IR-optical log slope of –0.75, and an SMC extinction curve, typical for extinction within other galaxies.)

Compton telescopes, described elsewhere (e.g., [28]) are typically used for detection in the 200 keV to ~ MeV energy range. In addition to providing high-energy detection, they can provide polarization information via reconstructing the Compton scatter interaction, which is polarization dependent. We have studied the addition of an instrument like GROME/GRIPS[29,30], which provides both of these capabilities; such an instrument requires a substantial additional mass and power allotment, but provides great benefit. First, polarization information helps to establish (or refute) a synchrotron origin of the emission (Section 2.a). Second, high-energy capability will increase the fraction of short-type GRBs, a type detected much less frequently by *Swift* or other instruments without > 200 keV response (10% for *Swift* [27] vs. ~20% for higher energy instruments[31], noting this value can be sensitivity dependent). High-energy capability also helps determine the peak energy, an important parameter used in correlation studies (Section 2.b). This would be a valuable, desirable option for a larger mass-budget mission.

For the UFFO-100 configuration we have assumed power, some thermal regulation, data bus, transmission, and other services are provided by the spacecraft and not included in our mass numbers. We have assumed that our spacecraft is pointing-stabilized relative to fixed distant sky targets for ~ 60 s stretches. Our contingency for a spacecraft that does not provide this capability is to compensate for the motion of the spacecraft using the beam-steering system; we have not completed component selection for this contingency at this time, however. This contingency also requires precise pointing information from the spacecraft; if this information is not available, we would need an additional fast-read star field camera to control the beam steering system for image motion compensation.

### f. Performance Estimates

We estimate, by scaling our X-ray camera to *Swift* performance by detecting area and assuming a 70% duty cycle, that we will detect about 64 GRBs/year. (Polar or other orbits with many passages through high background regions, or missions with sun aspect or orientation constraints will have lower duty cycles.) Conservatively assuming no correlation between X-ray and optical/IR brightness, using the *Swift* UVOT detection rates and fluxes (i.e. *without* assuming the existence of additional GRB with fainter optical/IR emission), assuming a prompt optical-IR log slope of –0.75, and using the known extinguished GRB fraction[24], we estimate that we will detect about 29 GRB in NIR or optical bands. We emphasize that our optical-IR detection rate per X-ray detection will be significantly higher than *Swift*, even with the same NIR/optical telescope aperture size, because of superior detector quantum efficiency, and because our NIR camera will detect many extinguished bursts. We did not include in our estimate any adjustment for the possibility that rapid-response may catch the bursts at an earlier, brighter flux, which would increase our detection rate further. Most GRB are discovered while fading, consistent with this possibility.

## 5. SUMMARY

Based on the relevant science questions described in Section 2, the characteristics of GRB observing described in Section 3b, and the constraints of partner missions of opportunity, we have outlined a set of instruments for flight on a low-earth orbit spacecraft that can provide rapid-optical/NIR response to GRBs. Our estimated mass is 122 kg (not including "contingency mass" sometimes required by spaceflight agencies). The fundamental goal of the observatory is to provide a systematic study, with good statistics, of the first 60 s of GRB optical/IR emission, measurements beyond the abilities of *Swift* and other instruments. For these goals, we plan only an X-ray coded mask camera with CdZnTe semiconductor detectors for initial location of the GRBs, and a telescope with a beam-steering mechanism and optical and NIR cameras for rapid-response follow-up. For our X-ray instrument, with only 1024 cm$^2$ of detector area, significantly less than *Swift*, we loose a surprisingly small number of GRBs detectable by *Swift*, due to the unusual source count-fluence or -peak flux relation of GRBs. Even though our optical/NIR telescope aperture size is the same as *Swift*, our optical and NIR cameras are significantly more sensitive, due to superior quantum efficiency of more modern detectors. In addition, our NIR camera is sensitive in a spectral band where GRB optical emission is brighter, and where extinction is small, allowing the detection of many GRBs undetectable by *Swift*. Given these characteristics, our next generation observatory should produce 29 or more GRB detections in the NIR bands, more if some GRB are brighter at less than 60 s after trigger. Such observations would help our understanding of the mechanism of GRB emission, jet physics, the relation between optical, X-ray, and other GRB properties, and time-dependent evolution of dust extinction near the GRB source.

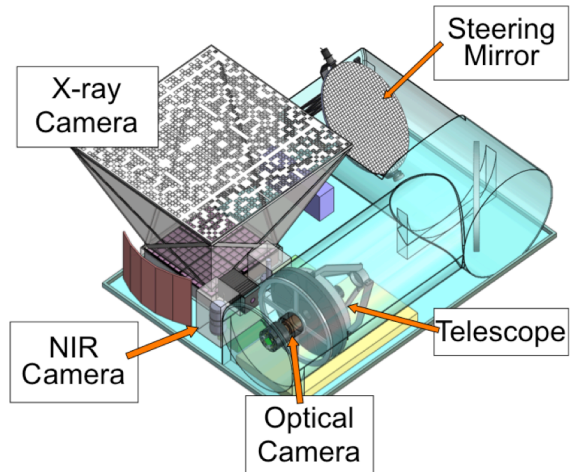

**Fig. 11** Rendering of the UFFO-100 (compact configuration) design. In this configuration, the overall envelope is minimized by use of a folding mirror.